\documentclass[11pt,a4wide]{article}

\usepackage{amsfonts}
\usepackage{amsbsy}
\usepackage{epsfig}
\usepackage{latexsym}
\input amssym.def
\input amssym.tex

\advance \topmargin by -\headheight
\advance \topmargin by -\headsep
\evensidemargin \oddsidemargin
\advance\hoffset by -3mm  % A4 is narrower.
\advance\voffset by  8mm  % A4 is taller.
\def\bbbc{{\mathchoice {\setbox0=\hbox{$\displaystyle\rm C$}\hbox{\hbox
to0pt{\kern0.4\wd0\vrule height0.9\ht0\hss}\box0}}
{\setbox0=\hbox{$\textstyle\rm C$}\hbox{\hbox
to0pt{\kern0.4\wd0\vrule height0.9\ht0\hss}\box0}}
{\setbox0=\hbox{$\scriptstyle\rm C$}\hbox{\hbox
to0pt{\kern0.4\wd0\vrule height0.9\ht0\hss}\box0}}
{\setbox0=\hbox{$\scriptscriptstyle\rm C$}\hbox{\hbox
to0pt{\kern0.4\wd0\vrule height0.9\ht0\hss}\box0}}}}

\makeatletter
\@addtoreset{equation}{section}
\makeatother

\newcommand\eqref[1]{(\ref{#1})}
\newcommand{\df}{{\mbox{\rm d}}}

\begin{document}
\title{{\Large \bf{Static Axisymmetric Einstein Equations in Vacuum: Symmetry, New Solutions and Ricci Solitons}}}
\author{
M M Akbar$^{1,}$\footnote{E-mail: akbar@utdallas.edu}
\ \,\,\&\,
M A H MacCallum$^{2,}$\footnote{E-mail: m.a.h.maccallum@qmul.ac.uk}
\\
\\ {$^{1}${Department of Mathematical Sciences}}
\\ {University of Texas at Dallas}
\\ {Richardson, TX 75080, USA}
\\
\\ $^{2}${School of Mathematical Sciences}
\\ {Queen Mary University of London}
\\ {London E1 4NS, UK}}
\date{\today}
 \maketitle
\begin{abstract}
  \noindent
  An explicit one-parameter Lie point symmetry of the four-dimensional vacuum Einstein equations with two commuting hypersurface-orthogonal
  Killing vector fields is presented. The parameter takes values over all of the real line and the action of the group can be effected
  algebraically on any solution of the system. This enables one to construct particular one-parameter extended families of axisymmetric static
  solutions and cylindrical gravitational wave solutions from old ones, in a simpler way than most solution-generation techniques, including the prescription given by Ernst for this system \cite{Ernst1978}. As examples, we obtain the families that generalize the Schwarzschild
  solution and the $C$-metric. These in effect superpose a Levi-Civita cylindrical solution on the seeds. Exploiting a correspondence between
  static solutions of Einstein's equations and Ricci solitons (self-similar solutions of the Ricci flow), this also enables us to
  construct new steady Ricci solitons.
  \end{abstract}
 \section{Introduction}
\label{intro}
The high nonlinearity of the Einstein equations makes them extremely difficult to solve. It makes it hard to draw generic physical conclusions about gravity and besets quantization. However, soon after Einstein found his equations, and thought them unsolvable, the first exact solution, describing the spacetime around a spherically symmetric massive object, was obtained by Schwarzschild. Since then Einstein's equations have been systematically studied for different matter fields subject to various local symmetries, algebraic conditions and other simplifying assumptions, and today we have many exact solutions in four dimensions that are well understood \cite{ExactSol2003, Griffiths, MacCallum2006}. These solutions provide concrete means to study the nonlinearities of the gravitational field. They shed light on more general non-exact solutions, guide numerical study and play a pivotal role in every quantum gravity program \cite{Bicak2000}.
Their study has brought the physics and mathematics communities together.

The difficulty of directly integrating Einstein's equations has led to many solution-generation techniques in which one obtains a solution, or a family of solutions, from a ``seed" solution, of the same system or a different system. In 1954 Buchdahl showed how to obtain a Ricci-flat solution from another in the presence of a hypersurface-orthogonal Killing
vector field \cite{Buchdahl1954} (see section \ref{Examples}). Ehlers in 1957 showed how one could obtain a stationary axisymmetric metric
starting from any static metric \cite{Ehlers}. Later, in 1972, Geroch showed that one can use the two commuting Killing vector fields of any stationary axisymmetric metric to obtain an infinite-parameter family of solutions \cite{Geroch1,Geroch2}. Following the discovery of Tomimatsu--Sato solutions \cite{Tomimatsu-Sato1,Tomimatsu-Sato2}, stationary axisymmetric systems were vigorously studied, aided by techniques developed in
other systems of partial differential equations (various B{\"{a}}cklund and other transformations, inverse-scattering methods
\cite{Belinski} etc.). Many sophisticated general results and specific solutions were obtained for stationary axisymmetric systems including
the Einstein--Maxwell system (see \cite{Hoenselaers:1985qk} and \cite[Chapter 34]{ExactSol2003}). However, applying those results to obtain
explicit solutions, of the same system or another, often involves solving an associated set of equations and performing a good number
of mathematical steps. One cannot usually simply write down a new solution starting from a seed solution.

Although the impressive work in four dimensions, and current efforts in obtaining higher-dimensional gravitational solutions modeled on the
four-dimensional ones, may suggest that there is little left to explore analytically for the four-dimensional Einstein equations with physically interesting symmetries and simple matter fields, in particular the vacuum, there is still more to be known. We present here one such unexpected new development.

We study the  vacuum Einstein equations in the presence of two commuting hypersurface-orthogonal Killing vector fields. In Lorentzian
four dimensions, these are axially symmetric static solutions and (Einstein--Rosen) cylindrical gravitational waves and can be obtained
from one another by a complexification of appropriate coordinates. In particular, we find a one-parameter Lie group that is a symmetry of
the system and maps any solution into a one-parameter extended family. In addition, the action of the group can be represented
algebraically. This produces, for example, an axially symmetric family that contains the spherically symmetric Schwarzschild metric as a
special case and another family that contains the $C$-metric (both families being distinct from the generalizations previously found).

The two systems -- the systems of vacuum static axisymmetric solutions and cylindrical wave solutions -- are well-studied in relativity. The
first gravitational wave solution found by Einstein and Rosen was cylindrical and the cylindrical wave system was among the very first
to be quantized \cite{Kuc71}. Despite the fact that cylindrical waves cannot describe radiation from an isolated body, they have been used to
understand energy loss due to gravity and the asymptotic structure of radiative spacetimes, test the quasilocal mass-energy of Thorne and in
cosmic censorship (see, for example, the review {\cite{Bicak2000}} and \cite{DiHerMac09}).

The present work was inspired by study of the Ricci flow equations, in particular the correspondences between Ricci solitons (self-similar
solutions of Ricci flow), the Einstein-scalar field theory and static vacuum solutions of the Einstein equations \cite{AkbarWoolgar}. The
symmetry of the axisymmetric vacuum system that we present here translates to an analogous symmetry for the corresponding steady Ricci
solitons, which we will discuss in section \ref{Ricci}. The results obtained are independent of the metric signature, and thus this work will be of interest to mathematicians looking at warped-product Ricci-flat metrics and warped-product Ricci solitons \cite{Besse, ONeill, Petersen, Petersen1}.
\section{The System(s)}
{\it{Static Vacuum System}}: It is well known that the general static axially symmetric vacuum solutions of Einstein's equations can be written in Weyl coordinates as
\begin{equation}
ds^2=-e^{2u(\rho, z)} dt^2+ e^{-2u(\rho, z)} \left[e^{2k(\rho, z)}(d\rho^2+dz^2)+\rho^2 d\phi^2\right]\label{basic1}
\end{equation}
where $u(\rho, z)$ and $k(\rho, z)$ satisfy the following three equations:
\begin{equation}
\frac{\partial^2 u}{\partial\rho^2}+\frac{1}{\rho}\frac{\partial u}{\partial\rho}+ \frac{\partial^2 u}{\partial z^2}=0,\label{laplace}
\end{equation}
\begin{equation}
\frac{\partial k}{\partial\rho}=\rho\left[\left(\frac{\partial u}{\partial\rho}\right)^2- \left(\frac{\partial u}{\partial z}\right)^2\right],\label{nonlinty1}
\end{equation}
\begin{equation}
\frac{\partial k}{\partial z}=2\rho \frac{\partial u}{\partial\rho}\frac{\partial u}{\partial z}\label{nonlinty2}.
\end{equation}
By ``a solution" we refer to a pair $(u, k)$ solving (\ref{laplace})-(\ref{nonlinty2}). The first equation (\ref{laplace}) is just the axially symmetric Laplace equation in cylindrical coordinates in an auxiliary three-dimensional Euclidean space. For any (harmonic) function $u (z, \rho)$ solving (\ref{laplace}),  $k(z, \rho)$ is uniquely determined and found by integrating (\ref{nonlinty1}) and (\ref{nonlinty2}), which reflect the nonlinearities of the Einstein equations. No distinction is made between solutions in which $u$ and/or $k$ differ by additive constants since they will give rise to the same metric by mere redefinitions of the coordinates. \\\\
{\it{Einstein-Rosen Cylindrical Wave System}}: It can be obtained from (\ref{laplace})-(\ref{nonlinty2}) by $z\rightarrow i\,t$ and $t\rightarrow i z$ and as such we will not separate it for discussion.
\section{Symmetries and Generating New Solutions from Old}
If $(u_1, k_1)$ and $(u_2, k_2)$ are two solutions, linearity of (\ref{laplace}) implies  $u=c_1 u_1+ c_2 u_2$ is a solution of (\ref{laplace}).
However, the nonlinearity of (\ref{nonlinty1}) and (\ref{nonlinty2}) prevents one from obtaining a standard prescription for $k$ in terms of
the four quantities $\{u_1, u_2, k_1, k_2\}$. One has to compute the line integral of (\ref{nonlinty1})-(\ref{nonlinty2}) (or some equivalent
set of differential equations) starting with $u=c_1 u_1+ c_2 u_2$, which is no different from the basic problem of solving (\ref{nonlinty1})-(\ref{nonlinty2}) for a given $u$. We discuss this general case further in section \ref{sec:group}.

Given an arbitrary solution $(u_0, k_0)$ can one generate another solution by some simpler means without solving the full set
(\ref{laplace})-(\ref{nonlinty2})? Ernst \cite{Ernst1978, Ernst1979} gave a method by which one can obtain a new solution  $(u_0+ cz,
k_0 +cF - \frac{c^2}{2}\rho^2)$ from a given solution $(u_0, k_0)$ provided the real function $F$ satisfies the following (simpler)
differential equation\footnote{In Ernst's paper, \eqref{ernst1} is misprinted (see \cite{Ernst1979}) and an auxiliary function $L$ is introduced which can be dispensed with. Unfortunately these oversights were repeated in \cite{ExactSol2003}, where $L$ was renamed $G$.}
\begin{equation}
\nabla F= 2i \rho \nabla u_0,\label{ernst1}\\
\end{equation}
where, in Weyl coordinates, $\nabla =\partial_\rho + i
\partial_z$.

Ernst's method superposes a multiple of the simple cylindrical solution with $u=z$ on $(u_0, k_0)$.  Ernst himself applied this to
obtain a generalization of the $C$-metric, and Kerns and Wild similarly obtained a one-parameter generalization of the Schwarzschild
metric \cite{KernsWildGRG}. For these one has to solve (\ref{ernst1}) starting with the seed's $u_0$, the difficulty of which depends on the
functional form of $u_0$.
\\
\\
{\it{Remark 3.1}} It is not necessary to transform the seed metric to Weyl coordinates in order to apply this transformation. Ernst notes that for
\begin{equation}
\df s^2= h[(\df
  x^1)^2+(\df x^2)^2]+\ell (\df x^3)^2-f(\df x^4)^2,\label{diagmet}
\end{equation}
one has the same
equation \eqref{ernst1} with $\nabla =\partial_{x^1} + i
\partial_{x^2}$ and $\rho^2=f\ell$. Also, the Weyl coordinates obey
\begin{equation}
\nabla z = i\nabla \rho.\label{zrhoeqn}
\end{equation}
 It is easy to generalize to
any coordinates in which the metric of $(x^1,\,x^2)$ space is diagonal: any overall factor in $\nabla$ can then be dropped as it
appears on both sides of the equations \eqref{zrhoeqn} and \eqref{ernst1}. This provides the simplest way of re-deriving the
new solutions given in \cite{Ernst1978, Ernst1979} and \cite{KernsWildGRG}.
\\
\\
Are there further ways of producing new solutions from old  without solving the field equations or an equivalent set of equations? One possible avenue that addresses this question is to look for explicit symmetries of the system. It is not difficult to see that the transformation
\begin{equation}
(u_0,k_0)\rightarrow (\beta u_0, \beta^2 k_0) \label{symm11}
\end{equation}
leaves the system (\ref{laplace})-(\ref{nonlinty2}) invariant; in other words, for any arbitrary solution $(u_0,k_0)$  there is a
(non-equivalent) solution $(\beta u_0, \beta^2 k_0)$ for $\beta \in(-\infty, \infty)$. More recently this has been used in
\cite{ChngMannStelea} to generate new solutions\footnote{The special case $(-u_0, k_0)$ will come up in section \ref{history}.}. However,
this transformation does not mix dependent and independent variables, which is why it was easy to find it by inspection. Below we present a
transformation that mixes variables in a nontrivial way. It is a parallel to Ernst's method in that it adds a multiple of a simple
cylindrical solution, in this case the solution $u=\ln \rho$ discussed below.  This prescription is clearly distinct from Ernst's, as we discuss further in section \ref{Examples}.
\\
\\
{\bf{Theorem 3.1}}: For $\alpha \in (-\infty, \infty)$, the transformation
\begin{equation}
(u_0, k_0) \rightarrow (u_0+\alpha \ln\rho, k_0+2\alpha u_0+\alpha^2\ln \rho),\label{symm2}
\end{equation}
leaves the system (\ref{laplace})-(\ref{nonlinty2}) invariant.
In other words, for every static axially symmetric vacuum solution of the Einstein equations
\begin{equation}
ds^2=\pm e^{2u_0(\rho, z)} dt^2+ e^{-2u_0(\rho, z)} \left[e^{2k_0(\rho, z)}(d\rho^2+dz^2)+\rho^2 d\phi^2\right]
\end{equation}
there exists a one-parameter generalization:
\begin{eqnarray}
ds^2&=&\pm e^{2u_0(\rho, z)} \rho^{2\alpha}dt^2+ e^{-2(1-2\alpha)u_0(\rho, z)}\rho^{2\alpha(\alpha-1)} \left[e^{2k_0(\rho, z)}(d\rho^2+dz^2)\right]\nonumber\\
&+&e^{-2u_0(\rho, z)} \rho^{2(1-\alpha)} d\phi^2.\label{transformedmetric}
\end{eqnarray}
\\
{\bf{Proof}}: By direct substitution of (\ref{transformedmetric}) into (\ref{laplace})-(\ref{nonlinty2}).\
\\
\\
Many papers in the literature speak in terms of ``Newtonian gravitational potentials" (which have no direct connection with the
actual Newtonian limit of the solution) since a solution $u$ of the Laplace equation (\ref{laplace}), which is of course the same as the
equation for an axisymmetric Newtonian gravitational potential in a vacuum, determines $k$ uniquely via
(\ref{nonlinty1})-(\ref{nonlinty2}). Applying the above transformation to $(u_0, k_0) \equiv (0,0)$, i.e.\ to (empty, flat) Minkowski space, we get
\begin{equation}
ds^2=-\rho^{2\alpha} dt^2+ \rho^{2\alpha^2-2\alpha}(d\rho^2 +dz^2) + \rho^{-2\alpha+2} d\phi^2\label{Levi-Civitamet}
\end{equation}
which is the Levi-Civita metric, one of the oldest and most widely used metrics in relativity (see \cite{MacCallumLeviCivita} for a
recent review). It is a particular case of the Kasner form, (13.51) in \cite{ExactSol2003}, which one can write as
\begin{equation}
ds^2= x^{2p} dx^2+x^{2a} dy^2+ x^{2b}
dz^2+x^{2c}dt^2, \label{Kasner}
\end{equation}
where the signature is in fact arbitrary and $a$, $b$, $c$ and $p$ satisfy the algebraic relations $a+b+c=p+1$ and $a^2+b^2+c^2=(p+1)^2$.
In terms of the Newtonian potential, therefore, what Theorem 3.1 is doing is superposing the Levi-Civita solution (\ref{Levi-Civitamet}) on the seed metric.
\\
\\
{\it{Remark 3.2.}} In the Riemannian (i.e.\ positive definite) signature, the transformation
\begin{eqnarray}
\alpha &\rightarrow& 1-\alpha \label{discretetrans1} \\
u &\rightarrow& -u\label{discretetrans2}
\end{eqnarray}
only interchanges the role of $\phi$ and $t$ in (\ref{transformedmetric}). These two geometries would therefore be indistinguishable locally.

\subsection{Group Properties, Symmetry and Solution Space}\label{sec:group}
To appreciate the special nature of our transformation, we discuss obtaining parameter-dependent new solutions from old ones  by the
superposition $u=u_0+\alpha {u_1}$ further.  One can imagine the whole space of solutions of  (\ref{laplace})-(\ref{nonlinty2}) being mapped into
itself under the influence of some ``external" field $\alpha {u_1}$, with $\alpha$ measuring its strength. The linearity of (\ref{laplace})
means $\alpha$ can take any value, so the resulting $(u,k)$ from $(u_0, k_0)$ would represent an infinite family of solutions -- a
curve in the space of solutions parametrized by $\alpha$ with $\alpha=0$ being the seed solution $(u_0, k_0)$. One would thus have a Lie
point symmetry of (\ref{laplace})-(\ref{nonlinty2}) for any choice of the ``external" field $u_1$. However, to write down a metric one also
needs to know $k$ explicitly. For a fixed ${u_1}$ the $k$ corresponding to $u=u_0+\alpha {u_1}$ depends on the functional form and the derivatives of $u_0$ and requires integration of (\ref{nonlinty1})-(\ref{nonlinty2}). Trying this for some simple choices of $u_1$ one can see that the resulting $k$ does not generally depend on $u_0$ in a prescribed functional way. It appears Ernst proceeded by trying this for $u=u_0+\alpha z$ and noticed that the addition of $\alpha z$ to $u_0$ creates additive terms for $k_0$ that can be obtained via the simpler equation  (\ref{ernst1}). What we found in this paper is that if one instead takes the external field to be $u_1= \ln \rho$, one obtains an {\it{explicit}} algebraic prescription for $k$ without
having to solve any associated set of equations.

The explicitness of our transformation (\ref{symm2}) makes it easy to
check its group properties directly. Denoting our transformation by
$T_\alpha$, one can check closure, $T_{\alpha_2} \circ
T_{\alpha_1}=T_{\alpha_1+\alpha_2}$, since successive transformations
with $\alpha_1$ and $\alpha_2$ take $(u_0, k_0)$ to
$(u_0+(\alpha_1+\alpha_2) \ln\rho, k+2(\alpha_1+\alpha_2)
u_0+(\alpha_1+\alpha_2)^2\ln \rho)$. The seed metric is the solution
at the identity $\alpha=0$ (in fact any metric within the family can be taken to
be at $\alpha=0$) and the existence of the inverse is immediate with $[T_{\alpha}]^{-1}=T_{-\alpha}$.

We note here that the scale transformation (\ref{symm11}) also gives a Lie group (written multiplicatively) by restricting $\beta$ to
values in $\mathbb{R}-\{0\}$ with $T_{\beta_2} \circ T_{\beta_1}=T_{\beta_1\beta_2}$, $\beta=1$ as the identity and $[T_{\beta}]^{-1}=T_{1/\beta}$. With slightly more careful calculations, and without actually having to solve for $F$, it is possible to verify that the Ernst prescription, treated as a transformation $T_c$ acting on the seed $(u_0, k_0)$, is also a Lie group with $c \in \mathbb{R}$ -- just like our transformation $T_\alpha$ above.

Contrasting with the closely related vacuum stationary system -- in which there exists a discrete map producing a new solution from an old
one  (cf. Eq (34.37) in \cite{ExactSol2003}) --  a one-parameter symmetry in the static vacuum case means the whole solution space of
the axisymmetric static vacuum Einstein system can be divided into equivalence classes of families that do not intersect under the action
of the group. One naturally wonders if there are other explicit transformations that could possibly connect these families. Note that
$z$ and $\ln\rho$ are the only one-variable functions possible here. Experimentation with other simple harmonic functions soon frustrates any hope of getting lucky. What is required is a systematic and careful symmetry analysis of the system; this is work in progress.

\subsection{Warped Form}
Despite the economical way Weyl coordinates express axially symmetric metrics, many physically and mathematically interesting solutions with
two commuting hypersurface-orthogonal Killing vector fields come in different coordinates and/or signatures, and may not possess axial
symmetry. The Schwarzschild metric, for example, despite having axial symmetry, is best described in its original spherical coordinates.
Interestingly, our symmetry (\ref{symm2}) can be rewritten as  transforming the general warped product
\begin{equation}
ds^2=\pm g_{11}(z^i) dx^2 \pm  g_{22}(z^i) dy^2+ g_{ij}(z^i) dz^i dz^j, \,\ i,j=3,4,
\end{equation}
with two line fibres corresponding to the two Killing vectors $\frac{\partial}{\partial x}$ and $\frac{\partial}{\partial y}$, in a nice way:
\\
\\
{\bf{Theorem 3.2}}: For every Ricci-flat metric of the form
\begin{equation}
ds^2=\pm g_{11}dx^2 \pm  g_{22} dy^2+ g_{ij}dz^i dz^j, \label{warpedformorg}
\end{equation}
where all metric components are functions of $z^i$ with $i,j=3,4$,
\begin{eqnarray}
ds^2= \pm (g_{22})^{\gamma} (g_{11})^{\gamma} g_{11}dx^2&\pm&(g_{22})^{-\gamma}(g_{11})^{-\gamma}g_{22}dy^2\nonumber\\
&+&(g_{22})^{\gamma(\gamma-1)}(g_{11})^{\gamma(\gamma+1)}g_{ij}dz^i dz^j, \label{warpedformorggen}
\end{eqnarray}
is Ricci-flat for $\gamma \in (-\infty, \infty)$.

One need not verify this by direct computation of the Ricci tensor of (\ref{warpedformorggen}) subject to the vanishing of the Ricci tensor
of (\ref{warpedformorg}) since this is just a rewrite of Theorem 3.1 with $\rho$ written as $\sqrt{g_{11}g_{22}}$ and $\alpha=\gamma$ (and other coordinates accordingly identified). (One can also view Theorem 3.2 as embodying the point made in Remark 3.1.) The advantage of working in this form is that one can write down the generalized metric without having to work out the $u$ and $k$ in Weyl coordinates. On the other hand, the Weyl form provides with the powerful, if sometimes misleading \cite{Bon92,ExactSol2003}, tool of considering solutions in terms of Newtonian potentials.

Note that the metric components $g_{ij}$ in (\ref{warpedformorg})-(\ref{warpedformorggen}) can assume arbitrary signatures; thus Theorem 3.2 can accommodate all possible semi-Riemannian metrics adapted to the two Killing vectors. The slightly elaborate form of the metric components in
(\ref{warpedformorggen}) is deliberate, to make the exponent structure manifest. Denoting by $T_{\gamma}$ the action that produces (\ref{warpedformorggen}) from (\ref{warpedformorg}), with very little algebra one can check that $T_{\gamma_2} \circ T_{\gamma_1}=T_{\gamma_1+\gamma_2}$, and $[T_{\gamma}]^{-1}=T_{-\gamma}$ etc. and verify the group properties of the transformation in these coordinates.
\section{Examples Extending Schwarzschild, C-metric and the Minkowski Metric}
\label{Examples}
We now apply our symmetry transformation to obtain some new exact solutions. There are plenty of other solutions, including cylindrical gravitational wave solutions, on which this can be applied equally easily and which we do not explore here.
\subsection*{The Schwarzschild Metric}
Applying $T_{\gamma}$ to the Schwarzschild metric
\begin{equation}
ds^2= -\left(1-\frac{2m}{r}\right)dt^2 + \frac{dr^2}{1-\frac{2m}{r}}+ r^2\left(d\theta^2 + \sin^2{\theta} d\phi^2\right), \label{SchOr}
\end{equation}
we obtain
\begin{eqnarray}
ds^2&=&-{r}^{2\gamma} \left(\sin\theta\right)^{2\gamma}\left(1-{\frac {2m}{r}}\right)^{\gamma+1} dt^2 \nonumber\\
&+& {r}^{2{\gamma}^{2}-2\gamma}\left(\sin\theta\right)^{2{\gamma}^{2}-2\gamma}\left(1-\frac {2m}{r}\right)^{{\gamma}^{2}+\gamma-1} dr^2 \nonumber\\
&+&{r}^{2{\gamma}^{2}-2\gamma+2} \left(\sin \theta\right) ^{2{\gamma}^{2}-2\gamma}
 \left( 1-{\frac {2m}{r}} \right) ^{{\gamma}^{2}+\gamma} d\theta^2 \nonumber\\
 &+& {r}^{2-2\gamma} \left( \sin\theta
 \right) ^{2-2\gamma} \left( 1-{\frac {2m}{r}} \right) ^{-\gamma} d\phi^2. \label{finalmet1}
\end{eqnarray}
This metric was not, as far as we know, written down before; it clearly has a more compact form than the generalization of Schwarzschild found by Kerns and Wild \cite{KernsWildGRG} using Ernst's prescription. Again, one could check that (\ref{finalmet1}) is indeed Ricci-flat by direct computation for $\gamma \in (-\infty, \infty)$. For $\gamma=0$ the spacetime symmetry group expands and one gets codimension-two spherical symmetry. There is obviously a number of ways to write (\ref{finalmet1}), including that the roles of $t$ and $\phi$ can be interchanged with simultaneous signature change etc.
\subsection*{The C-Metric}
We obtain the following generalization of the $C$-metric ($\gamma=0$ being the $C$-metric)
\begin{eqnarray}
ds^2= &-& \left( {\frac {-1+{y}^{2}-2\,m\, a\,{y}^{3}}{ \left( x+y \right) ^{2}}}
 \right) ^{\gamma+1} \left( {\frac {1-{x}^{2}-2\,m\, a\,{x}^{3}}{ \left( x+y
 \right) ^{2}}} \right) ^{\gamma} dt^2\nonumber\\
 &+&\left( {\frac {1-{x}^{2}-2\,m \,a\,{x}^{3}}{ \left( x+y \right) ^{2}}} \right) ^{\gamma \left( \gamma-1 \right) } \left( {\frac {-1+{y}^{2}-2\,m\, a\,{y}^{
3}}{ \left( x+y \right) ^{2}}} \right) ^{\gamma \left( \gamma +1 \right) }
 \left( 1-{x}^{2}-2\,m\, a\,{x}^{3} \right) ^{-1} \left( x+y \right) ^{-2}\,dx^2\nonumber\\
 &+& \left( {\frac {1-{x}^{2}-2\,m\, a\,{x}^{3}}{ \left( x+y \right) ^{2}}} \right) ^{\gamma\left( \gamma-1 \right) } \left( {\frac {-1+{y}^{2}-2\,m\, a\,{y}^{
3}}{ \left( x+y \right) ^{2}}} \right) ^{\gamma \left( \gamma+1 \right) }
 \left( -1+{y}^{2}-2\,m\,a\,{y}^{3} \right) ^{-1} \left( x+y \right) ^{-2}\,dy^2\nonumber\\
 &+&\left( {\frac {-1+{y}^{2}-2\,m\, a\,{y}^{3}}{ \left( x+y \right) ^{2}}} \right) ^{-\gamma} \left( {\frac {1-{x}^{2}-2\,m\, a\,{x}^{3}}{ \left( x+y
 \right) ^{2}}} \right) ^{-\gamma+1} \,dz^2.\label{CmetricGen}
\end{eqnarray}
This is clearly distinct from the generalized $C$-metric obtained by Ernst \cite{Ernst1978}.
\subsection*{The Minkowski Metric}

We could apply the transformation to the Minkowski metric in various coordinates. However, the result will just be a coordinate
transformation of the Levi-Civita metric \eqref{Levi-Civitamet}, which we obtained above. One can see this as follows. The coordinates must
give a form \eqref{warpedformorg}. One can transform from the standard Minkowski coordinates to the assumed form, apply Theorem 3.2, and then
reverse the coordinate transformation.

One might also hope that by applying Theorem 3.2 successively to two different choices of coordinates in which the metric has the form
\eqref{warpedformorg}, one could obtain a two-parameter solution. However, this fails because the first transformation will
give the form \eqref{Levi-Civitamet}. Then for any choice of coordinates in which the metric takes the form \eqref{warpedformorg},
  $g_{11}g_{22}$ will just be a function of the original $\rho$ and only a metric equivalent to \eqref{Levi-Civitamet} can result.

\subsection*{A Historical Link: Buchdahl's First Transformation}\label{history}
As was mentioned in the introduction, it was Hans Buchdahl who pioneered obtaining  new solutions from old ``without solving the
field equations''. In the 1950s \cite{Buchdahl1956, Buchdahl1959} he showed that if a Ricci-flat metric (i.e.\ a solution of the vacuum
Einstein equations) is ``static" in one of its coordinates one can obtain another distinct Ricci-flat metric from it by what he called a ``reciprocal transformation'' that takes the $d$-dimensional metric
\begin{equation}
ds^2= g_{ik}(x^{j})dx^i dx^k + g_{aa}(x^j) (dx^a)^2 \label{originalmetric}
\end{equation}
to the following $d$-dimensional metric
\begin{equation}
ds^2= (g_{aa})^{2/(d-3)}(x^{j})g_{ik}dx^i dx^k + (g_{aa})^{-1}(x^j) (dx^a)^2.\label{reciprocalmetric}
\end{equation}
Either metric, as Buchdahl termed, is ``$x^a$-static'', and it is easy to verify that if (\ref{originalmetric}) is Ricci-flat so is (\ref{reciprocalmetric}), by direct computation. By applying the transformation to (\ref{reciprocalmetric}) one gets back the original metric (\ref{originalmetric}).

It is easy to see that Buchdahl's reciprocal transformation in 4-dimensions is the $\alpha=\gamma=1$ case of Theorems 3.1 and 3.2 with the roles of $t$ and $\phi$ interchanged (cf.\ remark 3.2 with $\alpha=0$). It is also the $\beta=-1$ case of the scaling symmetry (\ref{symm11}) in Weyl coordinates.

{\it{Application:}} In his very first paper \cite{Buchdahl1956}, Buchdahl applied this transformation on
\begin{equation}
ds^2= -(dx^2+dy^2+dz^2)+ x^{2} dt^2,\label{firstBuchdahlSolutionsseed}
\end{equation}
and obtained the following Ricci-flat solution
\begin{equation}
ds^2= -x^4(dx^2+dy^2+dz^2)+ x^{-2} dt^2,\label{firstBuchdahlSolution}
\end{equation}
which is isometric to Taub's solution given as (15.29) in \cite{ExactSol2003}.
Applying his transformation to the Schwarzschild metric (\ref{SchOr}), static in its $t$ coordinate, he obtained
\begin{equation}
ds^2= -\frac{dt^2}{1-\frac{2m}{r}} + \left(1-\frac{2m}{r}\right) dr^2 + r^2 \left(1-\frac{2m}{r}\right)^2\left(d\theta^2 + \sin^2{\theta} d\phi^2\right),\label{SchT}
\end{equation}
which, upon the coordinate transformation $R=r-2m$, is again the Schwarzschild metric but with mass $-m$. Buchdahl expanded on the implication of his transformation for gravitational energy and showed that this is a special case of the component of a certain tensorial quantity, related to the Hamiltonian derivative of the Gaussian curvature, changing sign \cite{Buchdahl1956}. Buchdahl noted that more general solutions ``can be formed by means of a succession of reciprocal transformations, starting with the line element of a flat space". However, he did not apply this observation until much later \cite{Buchdahl1978}, in 1978, when he obtained from the all-positive version of (\ref{firstBuchdahlSolutionsseed}), i.e.\ from the flat space metric
\begin{equation}
ds^2= (dx^2+dy^2+dz^2)+ x^{2} dt^2,\label{firstBuchdahlSolutionsseed1}
\end{equation}
at the $(n-1)$th step of alternately taking the static coordinate $x^a$ to be $t$ and $z$ for the transformation, the solution
\begin{equation}
ds^2= x^{2n(n-1)}(dx^2+dy^2)+ x^{2n} dz^2+x^{-2(n-1)}dt^2. \label{secondBuchdahlSolution}
\end{equation}
This is readily recognized as again being of the Kasner form \eqref{Kasner} (and thus
Ricci-flat for all real $n$),
with $$
% p=a=\frac{n(n-1)}{n^2-n+1}, \quad b=\frac{n}{n^2-n+1,} \quad
% c=-\frac{n-1}{n^2-n+1}.$
p=a=n(n-1), b=n, c=-(n-1),
$$
and the same as \eqref{Levi-Civitamet} apart from signature.

As a second set of nontrivial Ricci-flat solutions, Buchdahl obtained from another form of the flat metric
\begin{equation}
ds^2= dx^2+dy^2+y^2 dz^2+ x^{2} dt^2\label{secondBuchdahlSolutionsseed}
\end{equation}
the following one-parameter family
\begin{equation}
ds^2= x^{2n(n-1)}y^{2(n-1)(n-2)} (dx^2+dy^2)+x^{2n} y^{2(n-1)} dz^2 +x^{-2(n-1)} y^{-2(n-2)} dt^2 \label{thirdBuchdahlSolutionKasnerform}
\end{equation}
which was known from the work of Harris and Zund \cite{Harris-Zund}. In summary, no new solutions were found by Buchdahl by this generation technique.

However, in all these calculations, what Buchdahl overlooked is that the Schwarzschild metric has another hypersurface-orthogonal Killing vector field, $\frac{\partial}{\partial \phi}$, which could be used to obtain a different Ricci-flat metric
\begin{equation}
ds^2= -r^4\sin^4{\theta} \left(1-\frac{2m}{r}\right)dt^2 + r^4\sin^4{\theta}\frac{dr^2}{1-\frac{2m}{r}}+ r^6\sin^4{\theta} d\theta^2 + \frac{1}{r^2\sin^2{\theta}} d\phi^2.\label{first}
\end{equation}
This would have been a new solution, which, unlike its $t$-counterpart (\ref{SchT}), is not related to the original Schwarzschild metric (\ref{SchOr}). Better yet, alternating between $t$ and $\phi$, as he did in his 1978 paper \cite{Buchdahl1978} to reproduce only the known  solutions (\ref{secondBuchdahlSolution}) and (\ref{thirdBuchdahlSolutionKasnerform}) from the flat metric, it is conceivable that Buchdahl could have arrived at our metric (\ref{finalmet1}) more than $30$ years ago. This would have provided him with a {\it bona fide} family of new solutions generalizing the Schwarzschild metric\footnote{If one applies Theorem 3.2 on (\ref{first}), one obtains the same family, differing from (\ref{finalmet1}) only in the sign of $m$ after changes of coordinates.}.

In  addition, and perhaps more importantly, Buchdahl did not give an explanation of why for two static coordinates the discrete exponents
produced by alternate transformations  also work fine for continuous values. Obviously, this question and its answer were hidden in the Lie point symmetry of the vacuum solutions of the Einstein equations with two commuting hypersurface-orthogonal Killing vector fields that we addressed here.
\section{Ricci Flow and Ricci Solitons}
\label{Ricci}
We now discuss a straightforward application of the above symmetry of the vacuum Einstein equations and produce new self-similar solutions of the Ricci flow with two commuting hypersurface-orthogonal Killing vector fields. For this we only review the basic definitions and readers are referred to standard references for more details \cite{CK, Chow1}.

Ricci flow is an intrinsic geometric flow in which the metric $g_{\mu\nu}$ on a manifold $M^{n+1}$ evolves by its Ricci curvature tensor
\begin{equation}
\frac{\partial g_{\mu\nu}}{\partial \eta} = -2R_{\mu\nu}
\label{eq1.1}
\end{equation}
along the flow parameter $\eta$, often referred to as ``time". It entered concurrently into the mathematics and physics communities
through the works of Richard Hamilton \cite{Hamilton82} and Dan Friedan \cite{Friedan} in the early 80s and has been used in
mathematics to study the interplay between geometry and topology of Riemannian manifolds. It was successfully applied to prove the long-standing Poincar\'{e} Conjecture and Thurston's Geometrization Conjecture (in three dimensions).

The simplest solutions of the Ricci flow are its fixed points
 \begin{equation}
\frac{\partial g_{\mu\nu}}{\partial \eta} = 0,
\end{equation}
which are the Ricci-flat metrics, $R_{\mu\nu}=0$.  The next simplest are the self-similar solutions in which the metric evolves only by rescalings and diffeomorphisms
\begin{equation}
g_{\mu\nu}(\eta)=\sigma(\eta) \psi^{*}_\eta (g_{\mu\nu}(0))\label{solmet}.
\end{equation}
It is easy to show that (\ref{solmet}) implies, and is implied by, the following equation for the initial metric (henceforth $g_{\mu\nu}$) \cite{CK}
\begin{equation}
R_{\mu\nu}-\frac12{\cal L}_X g_{\mu\nu}=\kappa g_{\mu\nu} \label{eq1.5}
\end{equation}
with $\sigma(\eta)=1+2\kappa\eta$ the scaling and $Y(\eta)=\frac{1}{\sigma(\eta)}X(x)$ the vector generating $\psi_\eta$ diffeormorphisms.

A Ricci soliton is a manifold-with-metric and a vector field $(M^{n+1},g_{\mu\nu}, X)$ solving (\ref{eq1.5}). The soliton is called
``steady" if $\kappa=0$, ``expander" if $\kappa<0$, and ``shrinker" if $\kappa>0$. A local Ricci soliton is one that solves (\ref{eq1.5}) on
an open region that might not cover a complete manifold with the soliton metric. A soliton is called gradient if $X=\nabla f$, where $f$ is a scalar function on $M^{n+1}$, and thus (\ref{eq1.5}) becomes
\begin{equation}
R_{\mu\nu}-\nabla_\mu \nabla_\nu f =\kappa g_{\mu\nu}. \label{grad}
\end{equation}
For $X=0$, or Killing, Ricci solitons (\ref{eq1.5}) are just Einstein metrics and hence trivial. The Cigar soliton, or Witten's black hole, is an example of a simple but nontrivial Ricci soliton, where
\begin{equation}
ds^2=\frac{dx^2+dy^2}{1+x^2+y^2},
\end{equation}
 and $X=2\left(x\frac{\partial}{\partial x}+ y\frac{\partial}{\partial y}\right)$.  It is a steady soliton on $\mathbb{R}^2$ solving (\ref{eq1.5}) with $\kappa=0$ and is gradient with $f=x^2+y^2$.
\subsection{Ricci Solitons and Static Metrics}
It is well-known \cite{Anderson2,AshtekarBicak, ExactSol2003} that if
\begin{equation}
ds^2 =\pm e^{2u}dt^2+e^{-\frac{2u}{n-2}}g_{ij}dx^idx^j\label{dRicciflat}
\end{equation}
is Ricci-flat in $(n+1)$-dimensions in which $\frac{\partial}{\partial t}$ is a hypersurface-orthogonal Killing vector field -- i.e.\ (\ref{dRicciflat}) is static in $t$ -- then $(u,g_{ij})$ solves the Einstein scalar field equations in $n$-dimensions
\begin{eqnarray}
&&R_{ij}-\frac{n-1}{n-2}\nabla_i u \nabla_j u
=0\ , \label{eq1.8}\\
&&\Delta u = 0 \ . \label{eq1.9}
\end{eqnarray}
 A precise relationship between Ricci solitons and Einstein-scalar field theory with a possible cosmological constant was given recently \cite{AkbarWoolgar} in which every solution of the latter in $n$-dimensions corresponds to a Ricci soliton in $(n+1)$-dimensions. In the case of zero cosmological constant this means every $(n+1)$-dimensional static vacuum solution (\ref{dRicciflat}) can be put in one-to-one correspondence with the following Ricci soliton metric in $(n+1)$-dimensions
\begin{equation}
ds^2 =e^{2\sqrt{\frac{n-1}{n-2}}u}dt^2+g_{ij}dx^idx^j\label{ndriccisol}
\end{equation}
with $X:=- 2\sqrt{\frac{n-1}{n-2}} g^{ij}\nabla_i u \frac{\partial}{\partial x^j}$. That steady solitons generated this way are necessarily incomplete in four dimensions follows from the inability of the Einstein-scalar system (\ref{eq1.8})-(\ref{eq1.9}) to admit any complete non-flat solution \cite{Anderson2}.

For any axisymmetric vacuum solution of the Einstein equations in Weyl coordinates
\begin{equation}
ds^2=\pm e^{2u(\rho, z)} dt^2+ e^{-2u(\rho, z)} \left[e^{2k(\rho, z)}(d\rho^2+dz^2)+\rho^2 d\phi^2\right]\label{basic1again}
\end{equation}
we therefore have the following local Ricci soliton
\begin{equation}
ds^2 =\pm e^{2\sqrt{2}u}dt^2+\left[e^{2k(\rho, z)}(d\rho^2+dz^2)+\rho^2 d\phi^2\right]\label{4driccisol}
\end{equation}
with $X=- 2\sqrt{2}e^{-2k(\rho, z)}\left(\nabla_{\rho} u \frac{\partial}{\partial \rho}+\nabla_{z} u \frac{\partial}{\partial z} \right)$.
\subsection{One-parameter Ricci Solitons}
Using the correspondence above and the Lie point symmetry (\ref{symm2}) we finally obtain the following one-parameter family of local  steady Ricci solitons
\begin{equation}
ds^2 =\pm e^{2\sqrt{2}u}\rho^{2\sqrt{2}\alpha}dt^2+ \left[e^{2k(\rho, z)+ 4\alpha u(\rho, z)} \rho^{2{\alpha}^2}(d\rho^2+dz^2)+\rho^2 d\phi^2\right]\label{4driccisolextended}
\end{equation}
with $X=- 2\sqrt{2}e^{-2k(\rho, z)+2\alpha u(\rho, z)+\alpha^2\ln\rho}\left(\frac{\alpha}{\rho}\frac{\partial}{\partial \rho}+ \nabla_{\rho} u \frac{\partial}{\partial \rho}+\nabla_{z} u \frac{\partial}{\partial z} \right)$ for every static axisymmetric vacuum solution of the Einstein equations (\ref{basic1}).
\section{Conclusion}
The primary motivation behind most solution-generation techniques has been to advance exact solutions, often starting from a particular
solution, and all require some form of integration. We found that the vacuum Einstein equations with two commuting hypersurface-orthogonal
Killing vector fields, which includes the axisymmetric system (\ref{laplace})-(\ref{nonlinty2}), admits a nontrivial exact Lie point
symmetry (\ref{symm2}) in explicit algebraic form. Being a symmetry of the system, this can be applied to generate one-parameter extended
families equally from known exact and non-exact solutions of the system (and thus can guide both analytical and numerical studies). The
new solutions can be seen as superposition of the seed metric with the Levi-Civita solution. The explicit nature of the
prescription means we do not have to solve any associated set of differential equations, and using it in the warped product form,
Theorem 3.2, means we do not have to convert to Weyl coordinates. This work interestingly connects to, and explains, some aspects of the very first generation technique given more than $60$ years ago -- and revisited from time to time for another two decades -- by Hans Buchdahl.

One can apply Theorem 3.1 or 3.2 to generalize any axisymmetric static or cylindrical gravitational wave solution and there is a  plethora of
possibilities. One can further combine this symmetry with the scaling symmetry (\ref{symm11}), and with Ernst's prescription \eqref{ernst1}, to write down more general multi-parameter families of metrics. In this paper, we limited ourselves to finding (new) generalizations of the Schwarzschild metric and the $C$-metric using this symmetry alone. As mentioned in the introduction, the initial motivation for looking into this well-studied system came from the recently found correspondence between Ricci flow and static metrics \cite{AkbarWoolgar}. The symmetry in the static system generalizes the corresponding Ricci solitons simultaneously.

One would naturally like to generate more solutions, study their properties, interpret and use them in relation to other known
solutions. However, the more important message that we believe comes from the existence of explicit symmetries like ours is that looking
vigorously and systematically for further hidden symmetries of the static system, and obtaining a  clearer picture of the geometry of the solution space, would be worthwhile\footnote{As a further motivation, we note here that symmetries (\ref{symm2}) and (\ref{symm11}) do not commute.}. A detailed and systematic study of symmetries would fall within the purview of the very developed field of symmetry analysis of nonlinear partial differential equations \cite{Olver, Stephani}. The related stationary system with two commuting vector fields, as we mentioned earlier, has been one of the most vigorously studied systems in relativity and may suggest methods, indicate symmetries, and help us understand the geometry of the solution space in general terms for the static case. Even though the static system is simpler, the connection is far less obvious. This is work in progress.

\section*{Acknowledgement}
We thank Behshid Kasmaie for useful discussions.

\end{document}